\documentclass{article}

\usepackage{arxiv}

\usepackage[utf8]{inputenc} 
\usepackage[T1]{fontenc}    
\usepackage{hyperref}       
\usepackage{url}            
\usepackage{booktabs}       
\usepackage{amsfonts,amsmath,amssymb, bm}       
\usepackage{nicefrac}       
\usepackage{microtype}      
\usepackage{cleveref}       
\usepackage{lipsum}         
\usepackage{graphicx}
\usepackage{natbib}
\usepackage{doi}
\newcommand{\hg}[1]{\hat g_{#1}}
\newcommand{\hG}[1]{\hat g^{#1}}
\newcommand{\hx}{\hat{x}}
\newcommand{\hy}{\hat{y}}
\newcommand{\ldot}{\dot{\mathcal L}}
\newcommand{\Lie}{\mathcal L}
\title{A fully observer-covariant formulation of the fluid dynamics of simple fluids: derivation, simple examples and a generalized Orr-Sommerfeld equation}

\newif\ifuniqueAffiliation
\uniqueAffiliationtrue

\ifuniqueAffiliation 
\author{ \href{https://orcid.org/0000-0001-8283-3070}{\includegraphics[scale=0.06]{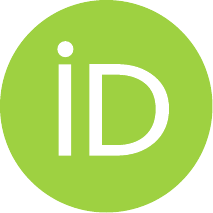}\hspace{1mm}Alberto  Scotti} \\
	School for the Engineering of Matter, Transport and Energy,\\
	Arizona State University\\
	Tempe, AZ \\
	\texttt{adscotti@asu.edu} \\
	}
\else
\usepackage{authblk}

\newbox{\orcid}\sbox{\orcid}{\includegraphics[scale=0.06]{orcid.pdf}} 
\author[1]{%
	\href{https://orcid.org/0000-0000-0000-0000}{\usebox{\orcid}\hspace{1mm}David S.~Hippocampus\thanks{\texttt{hippo@cs.cranberry-lemon.edu}}}%
}
\author[1,2]{%
	\href{https://orcid.org/0000-0000-0000-0000}{\usebox{\orcid}\hspace{1mm}Elias D.~Striatum\thanks{\texttt{stariate@ee.mount-sheikh.edu}}}%
}
\affil[1]{Department of Computer Science, Cranberry-Lemon University, Pittsburgh, PA 15213}
\affil[2]{Department of Electrical Engineering, Mount-Sheikh University, Santa Narimana, Levand}
\fi

\renewcommand{\shorttitle}{\textit{arXiv} Template}

\hypersetup{
pdftitle={A template for the arxiv style},
pdfsubject={q-bio.NC, q-bio.QM},
pdfauthor={David S.~Hippocampus, Elias D.~Striatum},
pdfkeywords={First keyword, Second keyword, More},
}

\begin{document}
\maketitle

\begin{abstract}
	We present a formalism to describe the motion of a fluid that is fully covariant with respect to arbitrary observers. To achieve full covariance, we write prognostic equations for quantities that belong to the graded exterior algebra of the cotangent bundle of the manifold occupied by the fluid. 
In particular, equations are that are fully covariant can be written for a purely Lagrangian observer, for which the fluid velocity (\textit{qua} section of the tangent bundle) is not a meaningful concept. 
With the new formalism, we consider problems of stability, and we derive a generalization of the Orr-Sommerfeld equation that describes the evolution of perturbations relative to an arbitrary observer. The latter is applied to cases where the observer is the Lagrangian observer comoving with the background flow.    
\end{abstract}


\section{Introduction}

In this work, we present a formalism for describing the dynamics of simple fluids that applies to arbitrary observers. Observers are grouped into two classes: Maxwellian and Leibnizian observers, with inertial (Galileian) observers constituting a subset of the Maxwellian observers.

Most textbook presentations of fluid mechanics more or less implicitly assume that the flow occupies a Riemannian manifold $\mathbb M$, and derive equations for the velocity field, an object that in modern parlance is a section of the tangent bundle of $\mathbb M$ \citep[see e.g.,][]{Thorne:2017}. 
In this framework, extending the Newtonian concept of force = change in momentum requires the introduction of a notion of parallel transport, which is based on the Levi-Civita connection, a.k.a. the covariant derivative.  However, this is not the only possible approach. Here, we develop a "quasi" forceless approach, where the quantities of interest are sections of the exterior algebra of the cotangent bundle. The approach can be derived from a Lagrangian point of view, and this naturally leads to considering a description in terms of Lie derivatives. 
 To illustrate this point, consider Kelvin's Theorem: It states that the circulation along a closed path as it evolves under the action of the flow in an ideal fluid can only be changed under the effect of a baroclinic torque (depending on the thermodynamics). A proper geometric formulation of Kelvin's theorem relates the change under the action of the flow of a closed 2-form, 
given by an extension of the Lie derivative to space-time, to the baroclinic torque. 

The equations of motion, initially derived for a certain class of observers, are extended to arbitrary observers applying the principle of covariance. 
At this point, the reader may wonder what the point is of all this. After all, the existence of a metric structure establishes a norm-preserving bijective correspondence between tangent and cotangent bundles (the so-called musical isomorphisms), and the covariant derivative based on the Levi-Civita connection commutes with the musical isomorphisms. 
However, as we shall see, for a purely Lagrangian observer, the "flow" is always zero, in the sense that if an observer sees something "pass by it" (which would be described by a path on the manifold whose derivative is a section of the tangent bundle), this would not be a Lagrangian observer! More precisely, the path of drifters for a Lagrangian observer is trivial: For a Lagrangian observer drifters always occupy the same position (that is, their coordinates do not change) on the manifold. Thus, the derivative of the trajectory is zero. Hence, the covariant derivative approach cannot be extended to a Lagrangian observer. Indeed, the approach in this case is to consider the accelerations relative to an external observer. The resulting equations of motions are then written in terms of the rate of change of the coordinates relative to a fixed frame and thus lack manifest covariance. In the present approach, the equations are manifestly covariant and apply as well to Eulerian and Lagrangian observers. The difference is that for an Eulerian observer, the metrical structure of the manifold is assigned \textit{a priori}, whereas for a purely Lagrangian observer the metrical structure emerges as part of the solution.

\section*{Background}

In this section, we provide a brief review of the relevant geometric concepts, definitions, and an initial formulation of the equations of motion that apply to a class of observers. In later sections, we will extend the equations to motion to cover arbitrary observers.  For a more detailed presentation, see \citet{scotti2016fluid}.
\subsection*{Basic geometric concepts}
\begin{itemize}
\item The fluid occupies a space described by a differentiable manifold $\mathbb M$ endowed with a Riemannian structure. In classical mechanics, time has an absolute character. The classical space-time  is $\mathbb G = \mathbb R \times \mathbb M$. A point $X\in \mathbb G$ is called an event. Notably, $\mathbb G$ is {\em not} endowed with a Riemannian structure. In other words, it is meaningless to measure the distance between two events that occur at different times. Lacking a metric structure in space-time, there is no natural way to map tangent to cotangent spaces of $\mathbb G$. 
\item We assume that the Riemannian manifold $\mathbb M$ is orientable and pick an orientation once and for all. Let the infinitesimal distance between two points whose coordinates are $(x^1+dx^1,\ldots,x^n+dx^n)$ and $(x^1,\ldots,x^n)$ be given by the quadratic form (the metric tensor)
$$
    dS^2=g_{ij}dx^i\otimes dx^j.
$$
$\otimes$ indicates the tensor product. 
The volume element in a coordinate patch is given by $\mathfrak{V}=\sqrt{|g|}dx^1\wedge\ldots\wedge dx^n$. Here, $\wedge$ denotes the exterior product, and $\sqrt{|g|}$ is the square root of the determinant of the metric tensor. When there is no risk of confusion, we will usually omit the symbol $\wedge$ when exterior multiplying differentials. 
\item In the exterior graded algebra of the cotangent bundle we have two fundamental linear operators which do not depend on the metric structure of the manifold; The exterior derivative $d$ and the interior product $i_{\bm v}$ by an element $\bm v$ of the dual space of the cotangent bundle (the tangent bundle). They are both antiderivations of the algebra. The exterior derivative maps $p-$forms to $(p+1)-$forms. On $0-$forms it is the standard differential. Furthermore, $d^2$ = 0. By linearity and the properties of antiderivation can be extended to any $p-$form.  
The exterior differential subsumes grad, curl, and div into a single albebraic structure. In particular we have the Stokes theorem
$$
\int_Sd\alpha = \int_{\partial S}\alpha, 
$$
where $\alpha$ is a $(p-1)-$form.  $S$ is a $p-$dimensional surface, and $\partial S$ is its boundary. 

The interior product maps $p-$forms to $(p-1)-$forms. On $1-$forms, 
$$
i_{\bm v}\alpha^1=\bm v(\alpha^1)
$$
(recall that $\bm v$ belongs to the dual space of $1-$ forms which is the space of linear functionals of $1-$forms) and on $0-$forms it sends scalar fields to 0.  
In addition, as required by antiderivations
$i_{\bm v} i_{\bm v}=0$. 
These definitions are, of course,
coordinate independent. 
Expressed in a local coordinate patch (using Eintein's repeated indices convention) 
$$
df=\frac{\partial f}{\partial x^j}dx^j\
$$
$$
i_{\bm v}\alpha=v^ji_{\bm \partial_j}a_mdx^m=v^ja_m\partial_j(dx^m)=v^ja_j.
$$

\item Although the formalism that we adopt applies to manifolds of arbitrary dimensions, in practice we are interested in two- and three-dimensional manifolds. In a coordinate patch, the natural (that is, holonomic) basis for $0-$ forms is $1$, and for $1-$ forms is, of course, $(dx^1,\ldots,dx^n)$.
For $(n-1)-$forms that are obtained taking the interior product of a $n-$form with a vector field $\bm v$, the natural basis is given by
$(\theta_1\equiv i_{\bm \partial_1}\mathfrak{V},\ldots,\theta_n\equiv i_{\bm \partial_n}\mathfrak{V})$. Finally, for $n-$ forms, the natural basis is simply the volume element. Note that when $n=2$ the $dx^i$ bases and the $\theta_i$ bases are not equivalent. In three-dimensional space, we have $\star (dx^p\wedge dx^j)=(|g|)^{-1/2}\star\epsilon^{ipj}\theta_i=(|g|)^{-1/2}\epsilon_i^{pj}dx^i$, where $\epsilon^{ipj}$ is the Levi-Civita symbol, and $\epsilon_i^{pj}=g_{il}\epsilon^{lpj}$.

\item The metric structure of the manifold induces an inner product (not to be confused with the interior product) on the $p-$forms. For our purposes, it is sufficient to define it on a basis. We have $\langle 1,1\rangle=1$, $\langle dx^i,dx^j\rangle=g^{ij}$, $\langle \theta_i,\theta_j\rangle=g_{ij}$ and $\langle\mathfrak{V},\mathfrak{V}\rangle=1$, where $g^{ij}g_{jp}=\delta^i_p$. 
\item The existence of a metric structure allows, via the Riesz representation theorem, the introduction of isomorphisms between the tangent and the cotangent space at a point $P$ on the manifold. The flat operator $\flat: TM_P\to T^*M_P$ is defined as 
$$
\bm v^\flat(\bm X)=\langle\bm v,\bm X\rangle,
$$
while its inverse $\sharp: T^*M_P\to TM_P$ 
$$
\alpha^\sharp(\beta)=\langle\alpha,\beta\rangle.
$$
These definitions are clearly coordinate-independent. 
In a local coordinate patch
$$
(\bm \partial_i)^\flat=g_{ij}dx^j,\,(dx^j)^\sharp=g^{ji}\bm \partial_i.
$$

\item The existence of an inner product on $p-$forms defines, via the Riesz representation theorem when the orientation is fixed, an isomorphism between $p-$ and $(n-p)-$forms, the Hodge star $\star$, fully defined by the relation 

$$\alpha\wedge \star\beta = \langle \alpha,\beta\rangle\mathfrak{V},    
$$
where $\alpha$ and $\beta$ are $p-$forms.
Again, for our purposes, it is sufficient to define it on the basis elements ($n$ is the dimension of the space). We have 
$$\star 1=\mathfrak{V},\,\star dx^j=g^{ji}\theta_i,\,\star\theta_i=(-1)^{n-1}g_{ij}dx^j,\star\mathfrak{V}=1.
$$
Note that for any $k-$form $\star\star\alpha=(-1)^{k(n-k)}\alpha$, where $n$ is the dimension of the manifold.  
\item The Lie derivative on forms. Let $\phi_s$ be a one-parameter family of diffemorphisms of $\mathbb M$ onto itself. Let $V(s)$ be the image under $\phi_s$ of a $p-$dimensional surface. Then
$$
\left.\frac{d}{ds}\int_{V(s)}\alpha \right|_{s=0}=\int_{V(0)}\Lie_{\bm v}\alpha,
$$
where $\Lie_{\bm v}$ is the Lie derivative and $\bm v$ at $P$ is tangent to $\phi_s(P)$.  
Cartan's magic formula expresses the Lie derivative in terms of interior product and exterior derivative as 
$$
\Lie_{\bm v}\alpha=d(i_{\bm v}\alpha)+i_{\bm v}(d\alpha). 
$$
The Lie derivative satisfies the Leibniz rule. Moreover, the Lie derivative commutes with the exterior derivative, but not with the interior product. 
\item Let $\alpha$ be a generic $p-$form. $\alpha$ is said to be closed if $d\alpha=0$. Conversely, if $\alpha=d\beta$ for some $(p-1)-$form, then $\alpha$ is called exact. Clearly, since $d^2=0$, exact forms are closed. Poincare's Lemma guarantees that if $\alpha$ is closed, then locally it is possible to find a $(p-1)-$form $\beta$ such that $\alpha=d\beta$. However, the global existence of $\beta$ is solely controlled by the topology of the manifold. In particular, if the manifold is simply connected (no holes), then closed forms are also globally exact.
\item A form $\alpha$ is said to be coclosed if $\star d\star\alpha=0$ and coexact if $\alpha=\star d\star\beta$. A coexact form is also coclosed. 
\item Hodge decomposition: on a closed Riemannian manifold, every form $\alpha$ is the orthogonal sum of a closed form plus a coclosed form plus a closed and coclosed form. The latter is called a harmonic form. In particular, if $\alpha$ is co-closed, then it is the sum of a coclosed form and a harmonic form. The restriction to closed Riemannian manifold is not too restrictive, since we always deal with finite energy cases, for which the manifold can be compactified.   
\end{itemize}
\subsection*{Observers and their relationships}
\begin{itemize}
\item Let $(x^0,x^1,\ldots,x^n)$ be the coordinates of the event $X$ for one observer in a given coordinate patch. A different observer will denote the same event with the coordinates 
$$
    \hx^0=x^0+s,
\,\hx^j=\hx^j(x^0,x^1,\ldots,x^n), j=1,\ldots,n.
$$

The fact that time between the two observers can only differ by an additive constant is required by the absolute character of time. No restrictions (other than smoothness as required) are placed on the $\hx^j$'s. Without loss of generality, we will always assume that all observers measure the same time (i.e., s=0). 
\item We can introduce an equivalence relation between observers. Two observers $\mathcal{A}$ and $\mathcal{B}$ are equivalent if the transformation that relates the spatial coordinates used by $\cal{A}$ to the coordinates used by $\cal{B}$ does not depend on time. In the following, an observer will always be understood modulo this equivalence relation.

\item Measurable quantities are encoded by elements of a suitable quotient space $G/T$ of the graded algebra 
$$
G\equiv\overset{n+1}{\underset{p=0}{\oplus}}\bigwedge^p\mathbb{G},
$$
where $n$ is the dimension of the manifold $\mathbb M$. The quotient is taken relative to the ideal 
$$
T=\{\alpha^p\in G: \forall s,p\,\, i_s^*\alpha^p=0\},
$$
where $\alpha^p\in \bigwedge^p\mathbb{G}$ is a $p-$form and 
$i_s$ is the one-parameter family of embedding maps\footnote{Not to be confused with the interior product $i_{\bm v}$ operator} 
$$
i_s: \mathbb{M}\to \mathbb{R}\times\mathbb{M}; i_s(P) \to (s,P).
$$
\item An admissible flow in space-time is the one-parameter Lie group whose generator, in a coordinate patch, is given by the vector field 
$$
    \bm v=\bm\partial_0+\sum_{i=1}^n v^i\bm\partial_i.
$$
Due to the restriction on time transformations between admissible observers, the 0th component of admissible  flows is always 1, regardless of the observer.  
\item If $\alpha^p$ is a $p-$form in the quotient ring $G/T$, the rate of change of $\int_S \alpha^p$, where $S$ is a $p-$dimensional surface, under the action of an admissible flow in space-time
is given by 
$$
    \int_S\left(\frac{\partial}{\partial t}+\Lie_{\bm v}\right)\alpha^p,
$$
where ${\Lie}_{\bm v}$ is the Lie derivative under the spatial component of the flow.
\item When applied to $0-$forms, the operator $\left(\frac{\partial}{\partial t}+\Lie_{\bm v}\right)$ coincides with the familiar material or Lagrangian derivative. 
\end{itemize}
\subsection*{Ideal observers measuring flows}
\begin{itemize}
\item An ideal observer seeds the flow with ideal 
instruments that measure thermodynamic quantities as well the rotation rate of the instrument. The instrument also records as a function of time its position $P$ on the manifold. The rotation rate of the instrument is expressed as a closed $2-$form $\Omega$. From the path of the instruments, a vector field is calculated $\bm v$, and we define for this observer  the one-form $\lambda=\bm v^\flat$.
Observers can fall into two categories: Maxwellian and Leibnizian observers. An observer is Maxwellian if $d\lambda=\Omega$. Otherwise, it is Leibnizian. Note that $\Omega$ is observer independent, in the sense that if $S$ maps one observer $\cal A$ to another observer $\cal B$, in $G/T$ $\Omega_{\cal A}=S^*\Omega_{\cal B}$. Not so $\lambda$!

\item Taking the thesis of Kelvin's theorem axiomatically and adding reasonable hypotheses (hydrostatic balance under no flow condition, the State Postulate of thermodynamics, conservation of energy in free systems) Maxwellian observers arrive at the following description of a perfect fluid 
\begin{subequations}
\begin{equation}
\left(\frac{\partial}{\partial t}+\Lie_{\bf v}\right)\lambda=d(E_k-{\mathcal Z})-\frac{dp}{\rho},\label{eq:lambda}
\end{equation}
\begin{equation}
\left(\frac{\partial}{\partial t}+\Lie_{\bf v}\right)\mathfrak{M}=0,
\label{eq:Mass}
\end{equation}
\end{subequations}
supplemented by an appropriate equation of state and by a transport equation for the second thermodynamic intensive quantity that, together with $p$, the pressure, specifies the density $\rho$. In the above equations, $\mathfrak M=\rho\mathfrak{V}$ is the mass $n-$form, expressed in terms of the density and the volume form $\mathfrak V$; $\mathcal Z$ is the potential per unit mass of the conservative force acting on the flow; $\bm v=\lambda^\sharp$ and $E_k=\langle\lambda,\lambda\rangle/2$ is the kinetic energy; and finally, $d$ is the exterior derivative restricted to the spatial part.    
\item For brevity, we introduce the notation 
$$
\ldot_{\bm v}(\lambda)=\left(\frac{\partial\lambda}{\partial t}+\Lie_{\bm v}\right)\lambda. 
$$
\item We can think of $\lambda$ as the (local if necessary) potential associated with the 2-form vorticity $\Omega=d\lambda$ whose existence is guaranteed by Poincare's lemma. In our formalism, it is this local vorticity potential, a section of the cotangent bundle, which is prognosed. Moreover, the vorticity (\textit{qua} 2-form) plays a more fundamental role than momentum. Momentum density (understood as $\rho\bm v$, a section on the tangent bundle) is not an invariant quantity across observers. On the other hand, the total vorticity is, as will become clear later.    
\item Note that these equations are prognostic equations for objects that live in the graded exterior algebra of the cotangent bundle of $\mathbb M$, and that the "derivative" operator is given by the Lie derivative. We do not need the Levi-Civita connection (a.k.a., the covariant derivative).

\end{itemize}
\section*{From Maxwellian to Leibnizian  frames}
Eq.(\ref{eq:lambda}-\ref{eq:Mass}) are invariant to changes of coordinates that do not involve time. In other words, if $\phi:\mathbb M\to\mathbb M$ is a diffeomorphism of the manifold onto itself, then the equations are invariant under the pullback $\phi^*$ provided that $\phi$ does not depend parametrically on time. In view of this, if two frames are related by such a $\phi$, for our purposes, they are one and the same observers, and that justifies our notion of equivalence between observers.

The $\lambda_\mathcal{A}$ measured by a Maxwellian observer  $\mathcal{A}$ is related to the $\lambda_\mathcal{B}$ measured by another Maxwellian observer $\mathcal{B}$ under the pullback $S^*$ from $\mathcal{B}$ to $\mathcal{A}$ 
via 
$$
S^*\lambda_{\mathcal{B}}-\lambda_{\mathcal{A}}=d\phi. 
$$
As is often the case in physics, we want to extend the validity of the equations to arbitrary frames by applying the principle of covariance. 
 
  In this case, we need to apply the exterior calculus machinery on $\mathbb{G}$ to forms that belong to $G/T$. 

Assume that in a coordinate patch the coordinates $\hx^j$'s of a Maxwellian observer ${\mathcal A}$ relate to the coordinates $x^j$'s of an arbitrary observer $\mathcal B$ via
\begin{equation}
x^j=x^j(t,\hx^1,\ldots,\hx^n),j=1,\ldots,n 
\end{equation}
and without loss of generality we assume that time is the same for both observers.
 Recall that $(\partial_t+\Lie_{\bm v})$ is the Lie derivative
in $\mathrm{G}/T$, measuring the rate of change of forms under the action of a flow. For a given coordinate system, the flow in $\mathbb{G}$ is described by its group operators $u^\alpha\bm {\partial/\partial x^\alpha}$, where $u^0=1$. The application of the chain rule gives
\begin{equation}
\hat u^\alpha\frac{\bm\partial}{\bm\partial \hx^\alpha}=\hat u^\alpha\frac{\partial x^\beta}{\partial \hx^\alpha}\bm{\frac{\partial}{\partial x^\beta}}= u^\beta\bm{\frac{\partial}{\partial x^\beta}},
\end{equation}
that is the components change as contravariant vectors (no surprises here, as they belong to the tangent space). Note  $u^0=1$ so ${\bf u}$ is an admissible flow for $\mathcal B$. 
At this point we define the null flow for observer ${\mathcal A}$ as the flow whose components $(1,\hat u^{*1},\ldots, \hat u^{*n})$ satisfy
\begin{equation}
\hat u^{*j}\frac{\partial x^i}{\partial \hx^j}=-\frac{\partial x^i}{\partial \hx^0}. \label{eq:Comp_vel}
\end{equation}
As can be easily verified, the null flow is the vector field tangent to the trajectories of drifters in ${\mathcal A}$ that for $\mathcal B$ corresponds to $(1,0,\ldots,0)$, so that the same drifters in ${\mathcal B}$ appear stationary\footnote{Alas, in classical space time we cannot slow down time!}. 
An interesting twist now occurs. Suppose that ${\mathcal A}$ measures a flow with components
\begin{equation}
\hat u^i=\hat u'^i+\hat u^{*i}.
\end{equation}
The same components for ${\mathcal B}$ are 
\begin{equation}
 u^i=u'^j\frac{\partial  x^i}{\partial \hx^j},\label{eq:infhat}
\end{equation}
which is not surprising, as (\ref{eq:Comp_vel}) is nothing but classical  law of composition of velocities. 
The form $\hat\lambda$ in ${\mathcal A}$ is  
\begin{equation}
\hat\lambda=\hat{\bm u}^\flat=(\hat u'^j+\hat u^{*j})\hat g_{ji}d\hx^i, 
\end{equation}
Modulo $T$, the action under the pullback of the transformation $S$ connecting the two observers becomes 
\begin{equation}
S^*\hat\lambda=\left(\hat u'_j\frac{\partial \hx^j}{\partial x^i}+\hat u^*_j\frac{\partial \hx^j}{\partial  x^i}\right)d x^i=\lambda+\lambda^f.
\label{eq:lambdaf}
\end{equation}
The twist is that while the components of the null flow vector are mapped to zero, $\lambda^f$ associated to the null flow is not zero under the action of the pullback modulo $T$! 
Wrapping up, under the pullback to ${\mathcal B}$, (\ref{eq:lambda}) becomes 
\begin{equation}
\ldot_{\bf{ v}}( \lambda)=-\ldot_{\bf{ v}}(\lambda^f)-\frac{dp}{\rho}-d\left({\mathcal Z}-\frac{1}{2}\langle\lambda+\lambda^f,\lambda+\lambda^f\rangle\right)
\end{equation}
and the components of the group operator ${\bf v}$ are given by (\ref{eq:infhat}). After simple manipulations 
we arrive at the final form of (\ref{eq:lambda}) under the pullback (in space-time!)
\begin{equation}
\ldot_{\bf{v}}(\lambda)=\left[-\frac{\partial}{\partial t}\lambda^f-i_{\bf  v}d\lambda^f\right]-\frac{dp}{\rho}+d\left(E_k+\frac{1}{2}\langle\lambda^f,\lambda^f\rangle-\hat{\mathcal Z}\right), E_k=\frac{1}{2}\langle\lambda,\lambda\rangle \label{eq:lambda_EGT}
\end{equation}

Comparing (\ref{eq:lambda}) with (\ref{eq:lambda_EGT}) shows that, in an arbitrary frame, the change in $\lambda$ by the flow measured in the arbitrary frame is due, in addition to the effect of material causes (external potentials)  and changes in the thermodynamic state, to the effect of the bracketed terms in (\ref{eq:lambda_EGT}) containing $\lambda^f$, which we call, for reasons that will soon become clear, the generalized Coriolis force. Furthermore, the external potential is modified by the "energy" of $\lambda^f$.

We call $\lambda^f$ the frame circulation and $d\lambda^f$ the frame vorticity. They are a property of the observer $\mathcal B$, not of the coordinates that $\mathcal B$ uses in its own frame. 
Eq.(\ref{eq:lambda_EGT}), complemented by the equation for the conservation of mass and the transport equation for the second thermodynamic quantity, which are already manifestly covariant, describe the flow from the point of view of an arbitrary observer.   
As noted above, for Maxwellian observers $\lambda^f=d\phi$, with $\phi$ an arbitrary $0-$form. Within Maxwellian observers, we can isolate Galileian observers, for whom $\phi$ is an affine transformation, which remains affine under a Galileian boost, and thus $d\phi$ is constant. For a Maxwellian observer, the overall Coriolis force is exact and thus can be absorbed in the definition of the gravitational potential. For Galileian observers, the only effect is to shift the potential by a constant, which is, of course, immaterial. The only way to distinguish between Maxwellian and Galileian observers is to know the local distribution of masses that generate the potential, from which the position of the center of mass can be determined. If the difference between the potential deduced by the mass distribution and the one deduced from an analysis of the motion is a constant, the observer is Galilean, translating uniformly relative to the center of mass. Otherwise, the observer is Maxwellian, who can accelerate relative to the center of mass. The acceleration does not need to be uniform in space or time, as long as $d\lambda^f=0$.    
 
Again, it is important to stress that the sum $(\lambda+\lambda^f)_{\cal A}=S^*((\lambda+\lambda^f)_{\cal B})$ is covariant among all observers. Not so for the velocity under pushforward. For a given flow, we have Galileian observers for which $\lambda\neq 0$ and $\lambda^f=0$ (up to a constant). At the opposite end of the spectrum, we have purely Lagrangian observers for whom $\lambda=0$ and $\lambda^f\neq 0$. 
Ultimately, the reason is rooted in Kelvin's theorem: If no kinematical effects can change the circulation on a closed curve, then $\Omega$ must be covariant under an arbitrary change of observers. From there, it is just a matter of algebra to derive the equations for the local potential $\lambda$.

\section*{Lagrangian frames}
A special example of a Leibnizian frame is the frame of the fluid itself, that is, the Lagrangian frame.

In this frame, obviously, the fluid appears motionless, though not necessarily steady (e.g., pressure can vary with time), since the metric will be in general time dependent. For a Lagrangian observer, the problem becomes determining $\lambda^f$ itself.  

 In the fluid frame the equation for the frame lambda becomes  
\begin{equation}
\frac{\partial\lambda^f}{\partial t}=-\frac{dp}{\rho}-d({\mathcal Z}-E^f_k),
\label{eq:lambdafEvolution}
\end{equation}
where $E^f_k=\langle\lambda^f,\lambda^f\rangle$ is the "kinetic energy" of the frame. For a Lagrangian observer, the rate of change of $d\lambda^f$ is given by 
\begin{equation}
    \frac{\partial d\lambda^f}{\partial t}=-\frac{dp d\rho}{\rho^2}.
\end{equation}

In the Lagrangian literature, $d\lambda^f$ is known as the Cauchy integral \citep{Bennett}. For a barotropic fluid, $-dp/\rho=-dg$ for some $0-$form $g$ and so the Cauchy integral is conserved. In this case, 
Hodge decomposing the frame lambda 
\begin{equation}
\lambda^f=dh+d^\star\Phi+\gamma,
\end{equation}
we have that the scalar potential $h$ satisfies
\begin{equation}
h_{,t}+g+{\cal Z}-E^f_k=0.
\end{equation}
This is the unsteady version of Bernoulli theorem. Remember that here $h$ is the scalar potential calculated by the Lagrangian frame. 
The potential $\Phi$ (a $2-form$) is the solution of the elliptic problem 
\begin{equation}
    \Delta \Phi = \Omega(t=0),
\end{equation}
where $\Delta$ is the deRham-Laplace operator,  $\Omega(t=0)=d\lambda^f$ at $t=0$ and the gauge is fixed so that $d\Phi=0$. Note that $\Phi $ will be time dependent, since the deRham-Laplace metric depends on the metric which, in the Lagrangian frame, is time dependent. 
For a barotropic fluid, the Cauchy integral is not conserved. From a Lagrangian point of view, the solution of the problem then requires the integration of the equation for $d\lambda^f$ in time. 

\subsection*{From Lagrangian back to Eulerian}
The Lagrangian frame is unique among the Leibnizian frames in that $\lambda^f$  is  not known \textit{a priori}, but arises from the dynamics itself. Suppose that a Lagrangian frame determines the solution, that is, determines $\lambda^f$.  
The inversion problem consists in determining the coordinate transformation to a frame which need not be Maxwellian, but whose metric structure is known. That is,  given the lambda of the Lagrangian frame $\lambda^f$, and a set of coordinates $\hx^j$ in a frame in terms of which the metric $\hat{g}_{ij}$ is known, we need to determine the coordinate transformations $\hx^j=\hx^j(t,x^1,\ldots,x^n)$.  

This is achieved by essentially "inverting"  eq. \ref{eq:lambdaf}. This is because the pullback of $\lambda^f$ to the frame encodes the trajectories of the Lagrangian tracers in the rigid frame. 
Thus, 
\begin{equation}
\frac{\partial \hx^i}{\partial t}{\hat{\bm\partial_i}}=(\lambda^f)^\sharp=(u^*_idx^i)^\sharp=\left(u^*_i\frac{\partial x^i}{\partial\hx^j}d\hx^j\right)^\sharp
\end{equation}
Flattening the last equation leads to the following $n$ coupled equations
\begin{equation}
    \frac{\partial\hx^p}{\partial t}\frac{\partial\hx^j}{\partial x^i}\hat{g}_{jp}=u^*_i=u^*_i(t=0)-\int_0^t\left[\frac{\partial}{\partial x^i}({\cal Z}-E_k^f)-\frac{1}{\rho}\frac{\partial p}{\partial x^i}\right]dt,\label{eq:metric1}
\end{equation}
  where the last equality comes from the integration of (\ref{eq:lambdafEvolution})  (When the hatted frame is a Cartesian this result was first obtained by Weber. See Lamb, Art 15). This system must be solved subject to the initial condition given by the map from the Lagrangian to the "known" frame coordinates at $t=0$. Once the transformation is known, it is possible to calculate the metric tensor in the Lagrangian frame via the usual transformation 
  \begin{equation}
\hat{g}_{pq}d\hx^pd\hx^q=\left(\hat{g}_{pq}\frac{\partial\hx^p}{\partial x^i}\frac{\partial\hx^q}{\partial x^j}\right)dx^idx^j=g_{ij}dx^idx^j\label{eq:gij}
  \end{equation}
Note that (\ref{eq:metric1}) cannot be considered a pure Lagrangian approach. The coordinates are indeed Lagrangian labels, but the trajectories are expressed in the hatted frame.   
\subsection*{ The geometry of purely Lagrangian frames} From the point of view of a Lagrangian observer, the notion of a "rigid" frame (i.e., a frame where the metric is time independent and known \textit{a priori}) may be completely alien. 
Unless the flow is trivial, for a Lagrangian observer the physics of the flow determines the relative distance of fixed-coordinate points at any given time and thus the metric. While for an observer for whom the metric is known \textit{a priori} the problem is to determine the flow relative to the known underlying geometry, for a Lagrangian observer, for whom there is no flow,  the problem is to determine the geometry of the space itself. That is, to achieve a purely Lagrangian description, we need to determine how the metric tensor $g$ evolves over time, without making any reference to another frame. To do so, we initially assume the existence of a "hatted" frame, with the goal of eliminating any reference to a "hatted" frame at the end. From eq.(\ref{eq:metric1}) we have  
\begin{equation}
\frac{\partial}{\partial x^j}\left(\frac{\partial\hx^p}{\partial t}\frac{\partial\hx^q}{\partial x^i}\hat{g}_{qp}\right)+\frac{\partial}{\partial x^i}\left(\frac{\partial\hx^p}{\partial t}\frac{\partial\hx^q}{\partial x^j}\hat{g}_{qp}\right)=\frac{\partial u^*_i}{\partial x^j}+\frac{\partial u^*_j}{\partial x^i}.
\label{eq:metric2}
\end{equation}
From the same equation 
\begin{equation}
\hat g_{pq}\frac{\partial\hx^p}{\partial t}=u^*_s\frac{\partial x^s}{\partial\hx^q},
\end{equation}
or, which is the same,
\begin{equation}
\frac{\partial\hx^s}{\partial t}=\hat g^{sm}u^*_n\frac{\partial x^n}{\partial \hx^m}.
\label{eq:hatT}
\end{equation}
The only assumption is that on the hatted frame the metric is time independent. 
The rate of change of the distance $dS^2$ between two infinitesimally close points according to the Lagrangian observer is
\begin{equation}
    \frac{d(dS^2)}{dt}=\frac{\partial g_{ij}}{\partial t}dx^i\otimes dx^j.
\end{equation}
For the "hatted" observer 
\begin{multline}
    \frac{d(dS^2)}{dt}=\frac{d}{dt}(\hg{pq}d\hx^p\otimes d\hx^q)=\frac{d}{dt}\left(\hg{pq}\frac{\partial\hx^p}{\partial x^i}\frac{\partial\hx^q}{\partial x^j}\right)dx^i\otimes dx^j=\\
    \left(\frac{\partial\hg{pq}}{\partial\hx^s}\frac{\partial\hx^s}{\partial t}\frac{\partial\hx^p}{\partial x^i}\frac{\partial\hx^q}{\partial x^j}+\hg{pq}\frac{\partial^2\hx^p}{\partial x^i\partial t}\frac{\partial\hx^q}{\partial x^j} + \hg{pq}\frac{\partial\hx^p}{\partial x^i}\frac{\partial^2\hx^q}{\partial x^j\partial t}\right)dx^i\otimes dx^j= \\
    \left(
    \frac{\partial\hg{pq}}{\partial\hx^s}\frac{\partial\hx^s}{\partial t}\frac{\partial\hx^p}{\partial x^i}\frac{\partial\hx^q}{\partial x^j}+
    \frac{\partial}{\partial x^i}\left(\hg{pq}\frac{\partial \hx^p}{\partial t}\frac{\partial\hx^q}{\partial x^j}\right)-
    \frac{\partial\hx^p}{\partial t}\frac{\partial}{\partial x^i}\left(\hg{pq}\frac{\partial \hx^q}{\partial x^j}\right)\right.+\\
    \left.\frac{\partial}{\partial x^j}\left(\hg{pq}\frac{\partial \hx^q}{\partial t}\frac{\partial\hx^p}{\partial x^i}\right)-
    \frac{\partial\hx^p}{\partial t}\frac{\partial}{\partial x^j}\left(\hg{pq}\frac{\partial\hx^q}{\partial x^i}\right)
    \right)dx^i\otimes dx^j=\\
    \left(\frac{\partial\hg{pq}}{\partial\hx^s}\frac{\partial\hx^s}{\partial t}\frac{\partial\hx^p}{\partial x^i}\frac{\partial\hx^q}{\partial x^j}-
    \frac{\partial\hx^p}{\partial t}\frac{\partial}{\partial x^i}\left(\hg{pq}\frac{\partial \hx^q}{\partial x^j}\right)-
    \frac{\partial\hx^p}{\partial t}\frac{\partial}{\partial x^j}\left(\hg{pq}\frac{\partial\hx^q}{\partial x^i}\right)+
    \frac{\partial u^*_i}{\partial x^j}+\frac{\partial u^*_j}{\partial x^i}
    \right)dx^i\otimes dx^j = \\
    \left(
    \frac{\partial u^*_i}{\partial x^j}+\frac{\partial u^*_j}{\partial x^i}-
    \frac{\partial \hx^s}{\partial t}\left(
    2\hg{sq}\frac{\partial^2\hx^q}{\partial x^i\partial x^j}+
    \frac{\partial\hg{sq}}{\partial\hx^n}\frac{\partial\hx^n}{\partial x^i}\frac{\partial\hx^q}{\partial x^j}+
     \frac{\partial\hg{sq}}{\partial\hx^n}\frac{\partial\hx^n}{\partial x^j}\frac{\partial\hx^q}{\partial x^i}-
    \frac{\partial\hg{pq}}{\partial\hx^s}\frac{\partial\hx^p}{\partial x^i}\frac{\partial\hx^q}{\partial x^j}
    \right)
    \right)dx^i\otimes dx^j= \\
    \left(
    \frac{\partial u^*_i}{\partial x^j}+\frac{\partial u^*_j}{\partial x^i}-
    2u^*_n\left(\frac{\partial x^n}{\partial\hx^q}\frac{\partial^2\hx^q}{\partial x^i\partial x^j}+
    \frac{1}{2}\hG{sm}\frac{\partial \hg{sq}}{\partial\hx^p}\frac{\partial x^n}{\partial\hx^m}\frac{\partial\hx^p}{\partial x^i}\frac{\partial\hx^q}{\partial x^j}+
    \frac{1}{2}\hG{sm}\frac{\partial \hg{sq}}{\partial\hx^p}\frac{\partial x^n}{\partial\hx^m}\frac{\partial\hx^p}{\partial x^j}\frac{\partial\hx^q}{\partial x^i}-\right.\right.\\
    \left.\left.\frac{1}{2}\hG{sm}\frac{\partial \hg{pq}}{\partial\hx^s}\frac{\partial x^n}{\partial\hx^m}\frac{\partial\hx^p}{\partial x^i}\frac{\partial\hx^q}{\partial x^j}
    \right)
    \right)dx^i\otimes dx^j=\\
    \left(
    \frac{\partial u^*_i}{\partial x^j}+\frac{\partial u^*_j}{\partial x^i}-
    2u^*_n\left(\frac{\partial x^n}{\partial\hx^q}\frac{\partial^2\hx^q}{\partial x^i\partial x^j}+
    \frac{1}{2}\hG{sm}\left(
    \frac{\partial\hg{sq}}{\partial\hx^p}+
    \frac{\partial\hg{sp}}{\partial\hx^q}-
    \frac{\partial\hg{pq}}{\partial\hx^s}
    \right)
    \frac{\partial x^n}{\partial\hx^m}\frac{\partial\hx^p}{\partial x^i}\frac{\partial\hx^q}{\partial x^j}
    \right)
    \right)dx^i\otimes dx^j=\\
    \left(
    \frac{\partial u^*_i}{\partial x^j}+\frac{\partial u^*_j}{\partial x^i}-
    2u^*_n\left(\frac{\partial x^n}{\partial\hx^q}\frac{\partial^2\hx^q}{\partial x^i\partial x^j}+
    \hat\Gamma^m_{pq}\frac{\partial x^n}{\partial\hx^m}\frac{\partial\hx^p}{\partial x^i}\frac{\partial\hx^q}{\partial x^j}
    \right)
    \right)dx^i\otimes dx^j=\\
    \left(%
    \frac{\partial u^*_i}{\partial x^j}+\frac{\partial u^*_j}{\partial x^i}-
    2u^*_n\Gamma^n_{ij}\right)
    dx^i\otimes dx^j=2\widetilde{\nabla_g\lambda^f},
\end{multline}
where $\hat\Gamma^n_{ij}$ and $\Gamma^n_{ij}$
are the Christoffel's symbols of the Levi-Civita connection in the "hatted" frame and the Lagrangian frame respectively (note that the Christoffel's symbols do not transform as a $(1,2)$ tensor, hence the terms containing second partial derivatives) and the $\widetilde{\,\cdot\,}$ represent the symmetric part of the tensor. 
  Since $d(dS^2)/dt$ is obviously invariant across observers, we have 
  \begin{equation}
      \frac{\partial g}{\partial t}=2\widetilde{\nabla_g \lambda^f}, \label{eq:dS2}
  \end{equation}
  where $\nabla_g$ is the Levi-Civita connection associated to the time dependent metric $g$ of the Lagrangian observer. Note that this is the only place where the LC connection is needed. Indeed, the LC arises naturally from the derivation. 
  Together with the equation for $\lambda^f$ and the continuity equation 
  \begin{equation}
      \frac{\partial}{\partial t}(\rho\mathfrak{V}_g)=0,\label{eq:cont}
  \end{equation}
  (\ref{eq:dS2}) completes the description of the coupled dynamics and geometry of the Lagrangian observer, in a way that does not make any reference to another frame. 
  Note that eq.(\ref{eq:dS2}) is consistent with the well-known fact that the shear tensor (in a "rigid" frame) measures the rate of change of the displacement of nearby fluid parcels.  
For an incompressible flow we have 
\begin{equation}
    0=\frac{\partial\mathfrak{V}_g}{\partial t}=\frac{1}{2}\mathrm{Tr}(\widetilde{\nabla\lambda^f})=d(i_{\lambda^{f\sharp}}\mathfrak{V}_g)=d(\star_g\lambda^f).
\end{equation}
A few remarks are in order: 
\begin{itemize}
\item For rigid frames, there is no need to introduce the Levi-Civita connection. Viscous effects can be accounted with the deRham-Laplace operator, defined in terms of exterior derivatives and Hodge stars. The Levi-Civita connection arises naturally only when considering how relative distances change in the Lagrangian frame. 
\item For an inviscid incompressible flow in its Lagrangian frame the dynamics reduces to a (deceptively) simple form 
\begin{subequations}
    \begin{equation}
    \frac{\partial g}{\partial t}=2\widetilde{\nabla_g\lambda^f},
    \end{equation}
    \begin{equation}
        \frac{\partial\lambda^f}{\partial t}=d\phi,
    \end{equation}
    \begin{equation}
        d(\star_g\lambda^f)=0.
    \end{equation}
\end{subequations}
Here, the subscript $_g$ reminds us that the operator depends on the time-evolving metric that couples the equations, since in order to update $\lambda^f$ one needs to calculate $\phi$, which in turn satisfies a Poisson equation with the metric-dependent deRham-Laplacian. 
\end{itemize}
\section*{Stability in quasi-Lagrangian frames, Boussinesq approximation: a generalized Orr-Sommerfeld equation}
We examine the stability of a background flow from the point of view of an observer in the Lagrangian frame of the background flow. 

For this, we linearize (\ref{eq:lambda_EGT}) assuming that the frame lambda and the geometry of the Lagrangian background frame are given. The Boussinesq approximation consists in replacing $-dp/\rho-d\mathcal{Z}$ with $-bdz$, where $b=-g(\rho-\rho_0)/\rho_0$ is the buoyancy of the fluid. 

The perturbation satisfies 
\begin{subequations}
    \begin{equation}
        \frac{\partial\lambda}{\partial t}=-i_{\lambda^\sharp}d\lambda^f-b'dz-dp' -\nu \Delta \lambda,
        \label{eq:masterPert}
    \end{equation}
    \begin{equation}
        \frac{\partial b}{\partial t}+\star \Lie_{\lambda^\sharp}(\overline{b}\mathfrak{V})=\frac{\partial b}{\partial t}+\star (d\overline{b}\wedge i_{\bm \lambda^\sharp}\mathfrak{V})=-\kappa \Delta b\label{eq:pertbprime}
    \end{equation}
    \begin{equation}
d(i_{\lambda^\sharp}\mathfrak{V})=d(\star\lambda)=0. \label{eq:MassCons}
    \end{equation}
\end{subequations}
Here $\Delta$ is the deRham-Laplace operator accounting for diffusion effects, $\overline{b}$ is the background buoyancy profile, $b$ is the buoyancy perturbation and $\lambda$ is the lambda of the perturbation. Note that the absence of the time derivative of the volume element is due to the fact that, for an incompressible background flow, in Lagrangian coordinates the volume element must be constant in time.  
Consider now the Hodge decomposition $\lambda=d^{\star}\Psi$,\footnote{The closed and harmonic components are assumed zero because we assume the perturbation to be incompressible and the manifold simply connected.} where $\Psi$ is a $2$ form. For simplicity, we consider a two-dimensional background flow, and two-dimensional disturbances.  Thus, $\Psi=\psi\mathfrak{V}$ for some scalar function $\psi$ and  
\begin{equation}
    d\lambda^f=\Omega\mathfrak{V}
\end{equation}
for some scalar function  $\Omega$.
Thus, the generalized Coriolis force becomes 
\begin{equation}
i_{\lambda^\sharp}d\lambda^f=\Omega i_{\lambda\sharp}\mathfrak{V}=\Omega(\star\lambda)=\Omega d\psi
\end{equation}
(recall that $\star dx^i=g^{ij}i_{\bm\partial_i}\mathfrak{V}$).

Taking the exterior derivative of (\ref{eq:masterPert})
\begin{equation}
    \frac{\partial d\lambda}{\partial t}=\frac{\partial (d d^\star\Psi)}{\partial t}=-d\Omega\wedge d\psi -db'\wedge dz-\nu \Delta (d d^\star\Psi).
    \label{eq:Orr-Somm}
\end{equation}
 which is an equation for $\psi$. The last equation with eq. \ref{eq:pertbprime} are the generalized Orr-Sommerfeld equations in the Lagrangian frame of the background wave. The elliptic deRham-Laplace operator has time-dependent coefficients, which are known because the Lagrangian geometry of the background flow is assumed to be known.   Finally, the pressure of the perturbation is determined by imposing (\ref{eq:MassCons}).

\section*{Examples}

In this section, we will consider examples of Leibnizian frames (we will use the terms observer and frame interchangeably).  The coordinates of an observer within a Galileian frame\footnote{This condition can be relaxed to be a Maxwellian frame.} will be "hatted", while "hat-free" variables will denote the coordinates of an observer in the non-inertial frame.
\ \\ {\bf Rotating layered frames}
Many geophysical flows are not far from a state of solid body rotation. In this case, it is expedient to study them from the point of view of a Leibnizian frame rotating with the fluid. Consider the  transformation connecting the cylindrical coordinates $(\hat x^1,\hat x^2,\hat x^3)$ of a Galileian frame to the coordinate $(x^1,x^2,x^3)$ of a Leibnizian frame
\begin{gather}
\hat{x}^1=x^1\,\mbox{(radial}\,\mbox{coordinate)},\\
\hat{x}^2=x^2+\omega t\,\mbox{(azimuthal}\,\mbox{coordinate)} ,\\
\hat{x}^3=\zeta(x^1,x^2,x^3,t)\,\mbox{(vertical}\,\mbox{coordinate)}.\label{eq:vertical}
\end{gather}
Relative to the Galileian frame, this Leibnizian frame rotates counter-clockwise (assuming $\omega>0$) around the $\hat x^3$ axis and measures the axial coordinate relative to a surface, which, in the Galileian frame, may not be stationary. Several types of coordinates can be represented this way, e.g. isopycnic coordinates if the surface is a surface of constant density or the so-called $\sigma-$ coordinates if the surface corresponds to a topographic boundary.
To calculate the frame lambda, we begin with the null velocity in the inertial frame, that is, the solution to 
\begin{equation}
\hat{u}^{*j}\frac{\partial x^i}{\partial \hat x^j}=-\frac {\partial x^i}{\partial t}. 
\end{equation}
Eq.~(\ref{eq:vertical}) gives $x^3$ implicitly via $\hat x^3-\zeta(\hat{x}^1,\hat{x}^2-\omega t,x^3,t)=0$, so that 
\begin{equation}
\frac{\partial x^3}{\partial\hat x^1}=-\frac{\zeta_{,1}}{\zeta_{,3}},\,\frac{\partial x^3}{\partial \hat x^2}=-\frac{\zeta_{,2}}{\zeta_{,3}},\,\frac{\partial x^3}{\partial \hat x^3}=\frac{1}{\zeta_{,3}},\,\frac{\partial x^3}{\partial t}=\omega\frac{\zeta_{,2}}{\zeta_{,3}}-\frac{\zeta_{,t}}{\zeta_{,3}},
\end{equation}
and  the null velocity is 
\begin{equation}
\hat u^{*1}=0,\,
\hat u^{*2}=\omega,\,
\hat u^{*3}=\zeta_{,t},
\end{equation}
where $\zeta_{,j}\equiv\partial\zeta/\partial \hat x^j$ and $\zeta_{,t}\equiv\partial\zeta/\partial t$.
The frame lambda in the Leibnizian frame is given by (we use the fact that the metric in the Galileian frame is diagonal with diagonal $(1,(\hat x^1)^2,1)$)
\begin{equation}
\lambda^f=\hat{u}^{*l}\hat{g}_{li}\frac{\partial\hat{x}^i}{\partial x^j}dx^j=
\omega (x^1)^2dx^2+\zeta_{,t}d\zeta.
\end{equation}  
 The frame vorticity is then
\begin{equation}
d\lambda^f=2\omega x^1dx^1dx^2+d\zeta_{,t}\wedge d\zeta\label{eq:frame_vort_iso}
\end{equation}
while the time derivative of the frame lambda
\begin{equation}
\frac{\partial\lambda^f}{\partial t}=d\left(\frac{(\zeta_{,t})^2}{2}\right)
+\zeta_{,tt}d\zeta
\end{equation}
Up to this point, we have not made any assumption on $\zeta(x^1,x^2,x^3,t)$ or the nature of $x^3$. In particular, if $\zeta_{,t}=0$, that is, the reference surface is stationary, 
then using (\ref{eq:frame_vort_iso}) the generalized Coriolis force
\begin{equation}
-i_{\mathbf{v}}d\lambda^f=2\omega x^1(v^2dx^1-v^1dx^2),\label{eq:frame_vort_Cor}
\end{equation}
coincides with  the standard expression of the Coriolis force.  
To write the equation for mass conservation, we begin by observing that the volume $3-$ form in the $(x^1,x^2,x^3)$ coordinates is
\begin{equation}
\mathfrak{V}=|\zeta_{,3}|x^1dx^1dx^2dx^3. 
\end{equation}
A useful choice of coordinates in large-scale geophysical flows is obtained when the density is used as an independent coordinate. 
With this choice, the 
 mass $3-$form becomes 
\begin{equation}
\mathfrak{M}=x^3\mathfrak{V}.
\end{equation}
 Let ${\bf v}=v^i\bm{\partial/\partial x^i}$. Mass conservation then requires
\begin{equation}
\begin{split}
\ldot_{\mathbf{v}}(\mathfrak{M})=x^3\ldot_{\bf v}(\mathfrak{V})+\mathfrak{V}\ldot_{\bf v}(x^3)=\\
[Q_{,t}+(x^1)^{-1}(x^1Qv^i)_{,i}]x^1dx^1dx^2dx^3=0.
\end{split}
\end{equation}
where $Q\equiv -\zeta_{,3}x^3$,  
which gives a prognostic equation for $Q$.\footnote{In order for the coordinate transformation to be well defined, $\zeta_{,3}\neq 0$, and thus it must be either positive or negative. In geophysical applications, density decreases with height, so that $|\zeta_{,3}|=-\zeta_{,3}.$} 

The heaving of the isopycnals, or, more generally, of the reference surfaces, introduces an extra term in the generalized Coriolis force.  
When the above equations are restricted to geophysical applications characterized by horizontal scales much larger than the vertical scale (thin layer approximation) and when the motion occurs on time scales much longer than $(\omega^{-1})$ (subinertial motion), it is possible to ignore most terms. In the equation for $\lambda$, the dominant balance for the component along $dx^3$ is hydrostatic, that is 
\begin{equation}
p_{,3}=-gx^3\zeta_{,3}=gQ,
\end{equation}
where $g$ is the gravitational acceleration and $p$ the pressure. Under the same approximations, $\zeta_{,tt}/g\ll 1$, therefore, the time derivative of the frame lambda can be ignored. Finally, as long as $v^3\zeta_{,3t}/g\ll 1$, and $\zeta_{,1}\zeta_{,2t}/\omega\simeq \zeta_{,2}\zeta_{,1t}/\omega\ll1$, the contribution of the heaving isopycnals to the frame vorticity can be neglected.
It is only when going beyond the thin layer approximation (e.g., going beyond the hydrostatic approximation, which necessarily implies moving toward suprainertial frequencies) that the heaving isopycnals contribution of the frame vorticity should be included.  

\ \\ {\bf The trochoidal wave}
Consider a Lagrangian frame related to a Cartesian Galileian frame by 
\begin{subequations}
    \begin{equation}
    \hx=x+e^{y}\sin(x+t),
    \label{eq:Trocha}
\end{equation}
\begin{equation}
    \hy=y-e^{y}\cos(x+t).
    \label{eq:Trochb}
\end{equation}

\end{subequations}
In the above, we work in a system of units where the basic units are speed and a wavenumber. 
To calculate $\partial x^\alpha/\partial\hx^\beta$ we notice that (\ref{eq:Trocha}-b) define $x^i$ implicitly.  To avoid cluttering the notation, define $S\equiv e^y\sin(x+t)$ and $C\equiv e^y\cos(x+t)$ Taking the differential 
\begin{subequations}
    \begin{equation}
        \left((1+C)\frac{\partial x}{\partial \hx}+S\frac{\partial y}{\partial\hx}-1\right)d\hx +\left((1+C)\frac{\partial x}{\partial \hy}+S\frac{\partial y}{\partial \hy}\right)d\hy +\left((1+C)\frac{\partial x}{\partial t}+S\frac{\partial y}{\partial t}+C\right)dt=0
    \end{equation}
    \begin{equation}
        \left((1-C)\frac{\partial y}{\partial\hx}+S\frac{\partial x}{\partial\hx}\right)d\hx+\left((1-C)\frac{\partial y}{\partial \hy}+S\frac{\partial x}{\partial \hy}-1\right)d\hy+\left((1-C)\frac{\partial y}{\partial t}+S\frac{\partial x}{\partial t}+S\right)dt=0,
    \end{equation}
\end{subequations}
we have a system of 6 equations in the six unknowns. Let define the matrix 
\begin{equation}
    A\equiv\left(
    \begin{array}{cc}
         1+C&S  \\
         S&1-C 
    \end{array}
    \right).
\end{equation}
The time derivatives satisfy 
\begin{equation}
    \left(\begin{array}{c}
         \frac{\partial x}{\partial t}  \\
         \frac{\partial y}{\partial t} 
    \end{array}
    \right)=-A^{-1}\left(\begin{array}{c}
         C  \\
         S 
         \end{array}
         \right).
\end{equation}
Similarly, 
\begin{equation}
    \left(\begin{array}{c}
         \frac{\partial x}{\partial \hx}  \\
         \frac{\partial y}{\partial \hx} 
    \end{array}
    \right)=A^{-1}\left(\begin{array}{c}
         1  \\
         0 
         \end{array}
         \right)
\end{equation}
and 
\begin{equation}
    \left(\begin{array}{c}
         \frac{\partial x}{\partial \hy}  \\
         \frac{\partial y}{\partial \hy} 
    \end{array}
    \right)=A^{-1}\left(\begin{array}{c}
         0  \\
         1 
         \end{array}
         \right),
\end{equation}
which combined give 
\begin{equation}
\frac{\partial x^i}{\partial \hx^j}=A^{-1}_{ij}.
\end{equation}
The null velocity in the Galileian frame satisfies
$u^{*j}{\partial x^i}/\partial \hx^j=-\partial x^i/\partial t$, whence the components in the Galileian frame
\begin{equation}
    (u^{*1},u^{*2})=(C,S)
\end{equation}
(Note that as Galileian components we understand $C=C(x(\hx,\hy),y(\hx,\hy))$ and similarly for $S$) The frame lambda in the Lagrangian frame can be easily calculated 
\begin{equation}
    \lambda^f=u^{*i}\hat{g}_{ij}d\hx^j=u^*_j\frac{\partial\hx^j}{\partial x^i}dx^i=dS+e^{2y}dx.
\end{equation}
As expected, the time dependence is contained in the closed component, and the frame vorticity is 
\begin{equation}
    d\lambda^f=-2e^{2y}dxdy=-2\sqrt{|A^{-1}|}e^{2y}d\hx d\hy=-2\frac{e^{2y}}{1-e^{2y}}d\hx d\hy.
\end{equation}
The kinetic energy of the frame (which is equivalent to the kinetic energy of the null velocity in the Cartesian frame) is $E_k^f=e^{2y}/2.$
As far as the dynamic is concerned, we have the Bernoulli equation (recall that $\partial S/\partial t=C$) 
\begin{equation}
{\rm const}=S_{,t}+P+{\cal Z}-E_k^f=C+P+g(y-C)-e^{2y}/2.    
\end{equation}
At the surface ($y=y_{max})$ the pressure is constant and uniform $P_0$. This requires that $g=1$, or in dimensional form $g=c^2k$, which provides the relationship between wavelength and celerity (note that technically this is not a dispersion relationship, since this is a nonlinear wave). Thus, 
\begin{equation}
    P=P_0+e^{2y}-e^{2^{y_{max}}}-(y-y_{max}).
\end{equation}
The amplitude of the wave is controlled by $y_{max}<0$. Note that in the limit $y_{\max}\to 0$ (maximal amplitude wave) the frame vorticity for the Lagrangian observers is finite, whereas for the Eulerian observer, the vorticity diverges.  It points to the singular nature of the crest. 
Finally, the metric in the fluid frame is 
\begin{equation}
    g_{ij}=\left(
    \begin{array}{cc}
         1+e^{2y}+2C& 2S  \\
         2S& 1+e^{2y}-2C 
    \end{array}\right),
\end{equation}
which can be used to calculate the distance between points and how it changes in time. 
This analysis can be used as a basis to calculate the stability of a wave. 

\ \\ {\bf 2D sheared frames}
By a 2D sheared frame we mean a 2-dimensional observer for whom a flow that for a Galileian observer consists of a 1D shear appears to be stationary. In other words,  the Cartesian coordinates of a Galileian frame are given by 
\begin{gather}
\hat{x}^1 = x^1 + tF(x^2),\\
\hat{x}^2 = x^2.
\label{eq:Coutte}
\end{gather}
From (\ref{eq:Comp_vel}) the components of the null velocity in the Galileian frame are $\hat{u}^{*1}=F(x^2),\,\hat{u}^{*2}=0$. The frame $\lambda^f$ is then easily computed
\begin{equation}
    \lambda^f=\hat{u}^{*j}\hat{g}_{ji}\frac{\partial\hat{x}^i}{\partial x^p}dx^p=F(x^2)(dx^1+tf(x^2)dx^2),
\end{equation}
where $f=F'$.
The metric in the sheared frame is time-dependent,
\begin{equation}
    g_{pq}=\hat{g}_{ij}\frac{\partial \hat{x}^i}{\partial x^p}\frac{\partial \hat{x}^j}{\partial x^q}, \,=> g_{11}=1,\,g_{12}=g_{21}=tf(x^2), \,g_{22}=1+t^2f^2(x^2).
    \label{eq:metricSheared}
\end{equation}
Notice that $\sqrt{g}=1$, as expected since the shearing flow conserves volumes and at $t=0$ the transformation is the identity, and thus the volume element in the sheared frame is simply $\mathfrak{V}=dx^1dx^2$. The geometry, as perceived by a sheared observer, is evolving in time, a very simple "expanding" universe. 

The frame lambda can expressed as  
\begin{equation}
    \lambda^f=d^\star\Phi, \,\Phi=\left(\int^{x^2}F(s)ds\right)dx^1dx^2,
\end{equation}
where the codifferential is defined as $d^\star=-\star d \star$ if the manifold dimension is even, while $d^\star=(-1)^k\star d\star$ where $k$ is the order of the form it acts upon if the dimension of the manifold is odd. Clearly, since $i_{\lambda^{f\sharp}}\lambda^f=F^2(x^2),$ we have 
\begin{equation}
    -\frac{\partial}{\partial t}\lambda^f+d\left(\frac{1}{2}i_{\lambda^{f\sharp}}\lambda^f\right)=0,
\end{equation}
thus the generalized Coriolis force in equation (\ref{eq:lambda_EGT}) reduces to \begin{equation}
    -i_{\bm v}d\lambda^f.
\end{equation}
In the special case of a linearly sheared frame ($f=S={\mathrm const.}$) aka a Couette frame, the Coriolis force
reduces to the standard Coriolis force of a frame rotating rigidly with angular velocity $S/2$. The only difference w.r.t. to a Couette frame is that in the latter the metric is time dependent.

\ \\ {\bf Internal waves} For a more interesting case, consider a stratified flow in the Boussinesq approximation, whereby $-dp/\rho-d{\cal Z}\simeq -(dp/\rho_0-bdz)$, where $z$ is local vertical, $b$ the buoyancy and $\rho_0$ a reference density. For simplicity, we consider a two-dimensional plane. Our inertial frame is a Cartesian frame rotated by an angle $\theta$ relative to the vertical. In other words, $z=\cos\theta \hat{x}^2-\sin\theta \hat{x}^1$ and the undisturbed buoyancy $b=N^2z=N^2(\cos\theta\hat{x}^2-\sin\theta\hat{x}^1)$ Our Lagrangian frame uses $x^1=b$ and $x^2=\hat{x}^2$ as coordinates (both of which are constant for a wave of appropriate frequency). That is we have 
\begin{gather}
    x^1=N^2(\cos\theta\hat{x}^2-\sin\theta(\hat{x}^1+Af(\hat{x}^2,t))),\\
    x^2=\hat{x}^2.
\end{gather}
which can be easily inverted to yield $\hat{x}^i=\hat{x}^i(x^1,x^2,t)$. Here $f(x,t)$ is an arbitrary function of its arguments. 
The null flow is 
\begin{equation}\hat{u}^{*1}=-A\partial_tf(\hat{x}^2,t),\, \hat{u}^{*2}=0.
\end{equation}
We thus have  
\begin{equation}
    \lambda^f={\hat u}^*_1\frac{\partial \hat{x}^1}{\partial x^i}dx^i=A\partial_tf(x^2, t)\left(\frac{dx^1}{N^2\sin\theta}-\left(\frac{\cos\theta}{\sin\theta}-A\partial_xf(x^2,t)\right)dx^2\right),
\end{equation}
form which we derive the frame vorticity 
\begin{equation}
d\lambda^f=-\frac{ A\partial^2_{xt}f(x^2,t)}{N^2\sin\theta}dx^1dx^2.
\end{equation}
Taking the exterior derivative of (\ref{eq:lambda_EGT}), and considering that in the Lagrangian frame $\hat{\bm v}=0$ and $dz=N^{-2}dx^1+A\sin\theta\partial_xfdx^2$, we arrive at the following equation 
\begin{equation}
    \frac{\partial^3f}{\partial x\partial t^2}=-(N\sin\theta)^2\frac{\partial f}{\partial x},
\end{equation}
whose general solution is 
\begin{equation}
    f(x,t)=F_1(x)\cos(\omega t)+F_2(x)\sin(\omega t), \omega^2=(N\sin\theta)^2,
\end{equation}
and we recover the well known dispersion relationship for internal waves. 
Incidentally, note that $\lambda^f$ is coclosed, meaning $d^\star\lambda^f=0$ (as can be expected, since the null flow conserves volumes), and indeed 
\begin{equation}
    \lambda^f=d^\star\left(A\int^{x^2}\partial_tf(s,t)ds\mathfrak{V}\right)
\end{equation}
Note that we have not made any assumptions regarding the magnitude of $A$.  Finally, to solve for the pressure, we solve the elliptic problem 
\begin{equation}
    (d^\star d+dd^\star)(p')=d^\star(Ax^1df),
\end{equation}
which gives the pressure $p=p'+(x^1)^2/(2N^2)+\langle\lambda^f,\lambda^f\rangle/2$. Recall that the deRham-Laplacian operator is in Lagrangian coordinates, and thus its coefficients are time dependent. The components of the metric are 
\begin{equation}
    g^{11}=N^4(1-\sin2\theta A\partial_xf+(\sin\theta A\partial_xf)^2),\,g^{12}=N^2(\cos\theta-\sin\theta A\partial_xf), \,g^{22}=1,
\end{equation}
and $\sqrt{g}=(N^2\sin\theta)^{-1}$ (assuming $\sin\theta>0$).
Under the assumption of small amplitude, the terms containing $A$ in the Laplacian can be neglected, resulting in the usual expression of the pressure for linear internal waves. 


\ \\ {\bf Stability of Couette flows}
The stability of Couette flows under infinitesimal perturbations, first considered by \citep{Hopf14}, was not definitively settled until much later, when \citet{Romanov73} proved that the Orr-Sommerfield equation yields stable eigenvalues at all Reynolds numbers \citep[see][and references therein]{DrazinReid}. Here, we show that in the Couette frame, the stability can be proven rather easily, resting fundamentally on the fact that in the Couette frame there is no source of vorticity for the perturbation. For simplicity, we choose a system of units based on shear and viscosity, and we choose the shear $S$ of the Couette flow as a unit of shear and the viscosity $\nu$ of the fluid as unit of viscosity. In these units, lengths have dimension $\sqrt{\nu/S}$. Stability is controlled by specializing the generalized  Orr-Sommerfeld equation (Eq.~\ref{eq:Orr-Somm}) to this particular case.  We set $\Omega=S=$const., $\overline{b}=b'=0$ and we use the metric derived earlier. we obtain

\begin{equation}
    \frac{\partial (dd^\star\Psi)}{\partial t}=-\Delta (dd^\star\Psi).
    \label{eq:FHeat}
\end{equation}
Thus $dd^\star\Psi$, which is the vorticity of the perturbation, satisfies a heat equation where the elliptic operator is time dependent. 
Recall that while the commutator $[d,\partial/\partial t]=0$, the commutator $[d^\star,\partial/\partial t]\neq 0$ in general if the metric is time dependent. In particular, the commutator applied to $1-$forms (in 2D)
\begin{equation}
    [d^\star,\partial/\partial t]=\star d \star_{,t},
\end{equation}
where $\star_{,t}$ indicates that the star operator is defined in terms of time derivative of the metric. With this in mind, the codifferential of (\ref{eq:masterPert})  provides a diagnostic equation for the pressure. 
In other words, the order of solution is as follows: \begin{enumerate}
    \item Solve (\ref{eq:FHeat}) for $\Omega=dd^\star\Psi$ subject to an initial condition for the vorticity $\Omega$. 
    \item Once $\Omega$ is known, solve the elliptic problem for $\Psi$, with appropriate boundary conditions 
    \item Finally, solve for pressure. 
\end{enumerate}
Note that in all of this the role of the Reynolds number is to set the size of the domain. Furthermore, in Step 2 $\Omega$ represents the source term. But since it satisfies a heat equation and we are in two dimensions, we can write $\Omega=\omega\mathfrak{V}$
$$
\star\Omega \wedge \frac{\partial\Omega}{\partial t}=\frac{1}{2}\frac{\partial\omega^2 }{\partial t} \mathfrak{V}=-\langle\Omega,dd^*\Omega\rangle\rangle\mathfrak{V}.
$$
Therefore, since $d^*$ is the adjoint of $d$, integrating over the manifold yields 
$$
\frac{\partial }{\partial t}||\omega||_2<0. 
$$
It is interesting to discuss the details of the decay. The metric is time dependent but uniform over the manifold (eq.~\ref{eq:metricSheared}). At $t=0$ we introduce in the flow a vortical disturbance with wavenumber $\bm k$ and amplitude $\omega_0$ is. Let $G(\bm k,t)=g(t)^{ij}k_ik_j$ be the principal symbol of the deRham-Laplace operator, which in this case, up to a sign, coincides with the standard Laplace-Beltrami operator (we have noted the explicit dependence of the metric on time). Then 
\begin{equation}
    \Omega(t)=\omega_0e^{ik_ix^i-\int_0^tG(\bm k,s)ds}\mathfrak{V}.
\end{equation}
Since $G(\bm k,t)=k^2(1-t\sin2\theta+t^2\cos^2\theta)$, the decay rate depends on the orientation of the wavenumber, but in general it will be $O(k^2)$, which means that a disturbance of size $O(L)$ will decay over a time $O(SL^2/\nu)$. 
In terms of the principal symbol, the energy density $E$ of the perturbation 
\begin{equation}
    2E=\langle \lambda,\lambda\rangle=\langle\mathfrak{F},dd^*\mathfrak{F}\rangle=
    \langle\mathfrak{F},\Omega\rangle=\omega_0^2\frac{e^{-\int_0^tG(\bm k,s)ds}}{G(\bm k,t)}.
\end{equation}
Depending on the orientation of the wavenumber, the energy density can exhibit a growing transient before the exponential decay sets in.  This is a reflection of the non normality of  the modes of the Orr-Sommerfeld equation in the Galileian frame.

\ \\ \subsection*{Localized perturbations of an internal wave beam}
The geometry in the Lagrangian frame of an internal wave beam was considered earlier. 
For simplicity, we consider a case in which the frame vorticity written as
$$
d\lambda^f=-i\omega S e^{i\xi}\mathfrak{V},
$$
where $S$ is a non dimensional parameter expressing the ratio of the amplitude of the along-beam oscillation to the across-beam wavelength, $\omega$ is the frequency of the background beam, and $\xi(x^2,t)$ is a phase function. We also make the assumption that $|S_{,x^2}|\ll |\xi_{,x^2}|$. In other words, the envelope of the beam contains many primary wavelengths. This condition can be relaxed, but as a first stab will do. 

We consider the evolution of a localized perturbation. We write the potential for the lambda of the perturbation as 
$$
\Psi=\psi\mathfrak{V},
$$
where $\psi$ is a function of the phase with dimensions of length squared divided by time. Furthermore, we assume that 
$$
\psi=\hat\psi(\xi)e^{ik_jx^j}
$$

We assume the wavenumber sufficiently high so that 
$$
dd^*\Psi\simeq k_jk_lg^{jl}(\xi)\hat\psi(\xi)e^{ik_jx^j}\mathfrak{V}=G(k,\xi)\hat\psi(\xi)e^{ik_jx^j}\mathfrak{V},
$$
and 
$$
d\psi\simeq i\hat\psi(\xi)e^{ik_jx^j}k_jdx^j.
$$
This is tantamount to assume that $|\xi_{x^2}||\ll |k|$. In a similar manner, we write 
$$
b'=\hat b(\xi)e^{ik_jx^j},
$$
and 
$$
db'=i\hat b(\xi)e^{ik_jx^j}k_jdx^j,
$$
with $\hat b(\xi)$ having a dimension of a length divided by a time squared. 
We now substitute this ansatz into the generalized Orr-Sommerfeld equation. We have $\partial/\partial t\to \omega d/d\xi$. Also, we replace $\hat\psi(\xi)\to \hat\psi(\xi)/G(k,\xi)$ (which now has dimensions of inverse of time)
$$
\omega\frac{d\hat\psi}{d\xi}=\left(iS\frac{N\omega^2\xi_{,x^2}k_1}{G(k,\xi)}e^{i\xi}-\nu G(k,\xi)\right)\hat\psi-i\frac{\omega}{N}(+Sk_1\omega Ne^{i\xi}-k_2)\hat b,
$$
$$
\omega\frac{d\hat b}{d\xi}=-i\frac{N\omega k_2}{G(k,\xi)}\hat\psi-\kappa G(k,\xi)\hat b.
$$
In matrix form
$$
\omega\frac{d}{d\xi}\left(\begin{array}{c}
     \hat \psi \\
     \hat b\end{array}\right)=\left(\begin{array}{cc
     }
          iS\frac{N\omega^2\xi_{,x^2}k_1}{G(k,\xi)}e^{i\xi}-\nu G(k,\xi) & -i\frac{\omega}{N}(+Sk_1\omega Ne^{i\xi}-k_2)  \\
          -i\frac{N\omega k_2}{G(k,\xi)} & -\kappa G(k,\xi)
     \end{array} \right)
     \left(\begin{array}{c}
     \hat \psi  \\
           \hat b
     \end{array}
     \right).
$$
To make the wavenumber dimensionally consistent, we replace $k_1\to N^{-2}k_1$ and let $(k_1,k_2)=k(\cos(\phi),\sin(\phi))$. To make the streamfunction and buoyancy perturbation dimensionally consistent we replace $\hat b\to \hat b N/k$. Now both $\hat\psi$ and $\hat b$ have dimensions of inverse of time.  The principal symbol becomes 
\begin{equation*}
\begin{split}
G(k,\xi)=k^2[(1-\cos^2\theta+(\cos\theta-S\sin\theta e^{i\xi})^2)\cos^2(\phi)+(\cos\theta-S\sin\theta e^{i\xi})\sin(2\phi)+\sin^2\phi]= \\k^2[(1-\cos^2\theta)\cos^2(\phi)+((\cos\theta-S\sin\theta e^{i\xi})\cos\phi+\sin\phi)^2]=k^2G(\theta,Se^{i\xi},\phi). 
\end{split}
\end{equation*}
We define a perturbation Reynolds as $Re=\omega/\nu k^2$, where $k$ is the perturbation wavenumber, and the Prandtl number $Pr=\nu/\kappa$. Finally, let $R=\xi_{x^2}/k$ be the dominant wavenumber of the beam nondimensionalized with the wavenumber of the perturbation. This gives us the final form of the stability problem 
$$
\frac{d}{d\xi}\left(\begin{array}{c}
     \hat \psi \\
     \hat b\end{array}\right)=
     \left(\begin{array}{cc}
          iSR\frac{\sin\theta\cos\phi}{G(\theta,Se^{i\xi},\phi)}e^{i\xi}-Re^{-1}G(\theta,Se^{i\xi},\phi) & -i(Se^{i\xi}\sin\theta\cos\phi-\sin\phi)  \\
          -i\frac{\sin\phi}{G(\theta,Se^{i\xi},\phi)} & -\frac{1}{Re Pr} G(\theta,Se^{i\xi},\phi)
     \end{array} \right)
     \left(\begin{array}{c}
     \hat \psi  \\
           \hat b
     \end{array}
     \right).
$$
The matrix has been completely nondimensionalized. It is periodic in $\xi$ with period $2\pi$. It is thus amenable to Floquet analysis.

\section{Conclusions}
In this paper, we have presented a fully covariant formulation of the equations of motion that describe the flow of simple fluids. Observers naturally fall into two categories: Maxwellian and Leibnizian observers. Maxwellian observers, which include Galileian observers as a subset, are defined as those observers for whom the vorticity (a 2-form) measured by ideal Lagrangian drifters is fully accounted for by the exterior derivative of the 1-form obtained by flattening the velocity field associated with the trajectories of the Lagrangian drifters. Leibnizian observers, on the contrary, observe a discrepancy between the measured vorticity and the calculated vorticity. The difference is a property of Leibnizian observers. Maxwellian observers form the most general class of observers for whom the Kelvin theorem applies. Extension of the equation of motions to a Leibnizian observer via the covariance principle introduces a generalized Coriolis force in the equations. With the introduction of the latter, the equations become fully covariant. 

To achieve full covariance, the prognosed quantities belong to the exterior graded algebra of the cotangent bundle of the manifold occupied by the fluid, and a suitable extension of the Lie derivative extends the familiar Lagrangian derivative to higher order forms. In general, there is no need to introduce a covariant derivative (that is, a Levi-Civita connection). 

We consider a few examples of Leibnizian observers, including rigidly rotating observers. The extension to more complex rotating observers, such as nutating observers, is trivial. 

A special case of a Leibnizian observer is a Lagrangian observer. In this case, there is no motion apparent, and the problem for such an observer is not to determine the velocity field (which is zero), but rather how the metric structure evolves in time. In this context, the Levi-Civita connection arises naturally. 

Finally, we derive a generalization of the Orr-Sommerfeld equation that describes the evolution of perturbations for observervers that are Lagrangian in the background flow and we apply to some canonical problems. 

\section{Acknowledgments}
I am indebted to my former students Dr. E. Santilli and Dr. S. Mendes. The former for suggesting that exterior calculus could be used to attack the problem and the latter for help with some of the technical points of exterior calculus.

\bibliographystyle{unsrtnat}
\bibliography{Localbib}  

\end{document}